# Strain evolution in GaN Nanowires: from free-surface objects to coalesced templates

M. Hugues[1*], P.A. Shields[2], F. Sacconi[3,4], M. Mexis[1], M. Auf der Maur,[3] M. Cooke[5], M. Dineen[5], A. Di Carlo[3], D.W.E. Allsopp[2], and J. Zúñiga-Pérez[1]

[1] CRHEA-CNRS, Rue Bernard Grégory, Valbonne, France
[2] Dept. of Electronic and Electrical Engineering, University of Bath, Bath, UK
[3] Department of Electronic Engineering, University of Rome "Tor Vergata", Rome, Italy
[4] Tiberlab Srl, Via del Politecnico, 1, 00133, Rome, Italy
[5] Oxford Instruments Plasma Technology, Yatton, Bristol, UK

*Corresponding author: mh@crhea.cnrs.fr

Top-down fabricated GaN nanowires, 250 nm in diameter and with various heights, have been used to experimentally determine the evolution of strain along the vertical direction of 1-dimensional objects. X-ray diffraction and photoluminescence techniques have been used to obtain the strain profile inside the nanowires from their base to their top facet for both initial compressive and tensile strains. The relaxation behaviors derived from optical and structural characterizations perfectly match the numerical results of calculations based on a continuous media approach. By monitoring the elastic relaxation enabled by the lateral free-surfaces, the height from which the nanowires can be considered strain-free has been estimated. Based on this result, NWs sufficiently high to be strain-free have been coalesced to form a continuous GaN layer. X-ray diffraction, photoluminescence and cathodoluminescence clearly show that despite the initial strain-free nanowires template, the final GaN layer is strained.

## I. INTRODUCTION

Devices architectures are becoming increasingly complex as they combine various kinds of heterostructures (e.g. quantum wells and quantum dots) and exploit specific properties related to the dimensions and geometry of particular objects, such as nanowires (NWs). The 1D geometry of the latter results in an enhancement of both light absorption and light extraction, making them optimum candidates for optoelectronic applications such as solar cells,[1,2] light-emitting diodes,[3] or lasers.[4,5] Indeed, thanks to rapid development during the 1990s, the first room-temperature lasing emission from NWs was reported as early as 2001 from a ZnO NW array,[4] followed only one year later by laser emission from a single GaN NW.[5] Since then, lasing emission from NWs has been demonstrated for many II-V and III-V materials. Moreover, the possibility to realize not only axial, but also core-shell, heterostructures has allowed the lasing emission wavelength to be tuned as "easily" as in conventional planar structures.[6,7]

The geometry of the NWs has also been advantageously used to fabricate sensors for chemical and biological species,[8,9] as well as for gas detection.[10] In comparison to their planar counterparts, their large surface-to-volume ratio promises very high detection sensitivity (more reaction area compared to volume) and improved response time (reduced diffusion length). In parallel to the huge development of NWs for optoelectronic and sensing applications,[11,12] their specific geometry has encouraged the study of new device functionalities based on controlled strain engineering and interesting piezoelectric properties (which are material dependent). For example, the ease of applying external stress to NWs and the large piezoelectric coefficients of large band-gap materials (ZnO and GaN), have been successfully combined to fabricate nanogenerators,[13,14] and very sensitive strain sensors.[15]

Efficient elastic relaxation of strain is another advantage of the NWs geometry. Indeed, almost strain-free NWs can be obtained even when grown on large lattice mismatched substrates owing to their large free-surface area. This is particularly interesting in the case of GaN, since the lack of native substrates results in large accumulated strains in epitaxial layers. The main reason is that the foreign substrates commonly employed (i.e. sapphire and silicon), have thermal expansion coefficients that differ from those of GaN. Thus, upon cooling down to room temperature after the growth, GaN layers are under compressive strain on sapphire and tensile strain on silicon. For the latter case, the huge tensile strain accumulated in the structures can lead to the appearance of cracks. To prevent cracking several strategies have been implemented, the most popular being the introduction of AlN interlayers.[16,17] This AlN provides a means of generating compressive strain during the growth in order to compensate the thermally-induced tensile strain generated during the cooling-down step. In this context, strain-free NWs have been proposed in 2001 as an alternative interlayer that would operate as a strain-release region.[18] Incidentally, since the NWs are also often defect-free, they could act as seeds for a modified version of epitaxial lateral overgrowth (ELO), a method widely developed fifteen years ago to significantly reduce the defect density of GaN layers.[19,20] Starting from a template made up of GaN NWs, the proposed method consists of coalescing the NWs until a continuous and flat layer is achieved. However, the strain state of the resulting layer remains unclear. Indeed, some authors have reported that the final coalesced layer is almost fully-relaxed,[18,21,22] while others groups have observed a strain rebuilt after coalescence.[23-25]

Before investigating the strain state of the final coalesced template, the strain state of the initial GaN NWs must be carefully characterized and the strain relaxation along the NW height must be known. This is important since it will allow the determination of the minimum NW height for which these nanostructures are strain-free and, thus, can be used as initial *passive* templates. Unfortunately, the determination of the strain evolution from the base of the NWs to their top is challenging compared to the characterization of the strain state inside NW heterostructures, which have been extensively studied over the past few years.[26-30] Two main techniques have been employed : synchrotron X-ray diffraction,[28,31] and high-resolution transmission electron microscopy (HRTEM).[26,27,29,30] The stability improvement of the synchrotron-based experiments have allowed Newton *et al.* to image the strain tensor of ZnO NWs in three dimensions.[32] However, the experiments have been carried out on single NWs separated from their initial substrate. Obviously this gives no information about the strain relaxation at the NWs base, which becomes a free-surface once separated from the substrate.[32] For the HRTEM studies, the NW diameters were small enough so that sample thinning was not required. However, in most cases the NWs were scraped off the substrate and dispersed onto a grid.[26,29,30] While some attempts have been carried out to study by HRTEM the strain at the NW-substrate interface, in this case sample thinning is unavoidable. Unfortunately, the strain state of small objects is certainly affected during the TEM preparation procedure, thereby altering the original strain distribution. Indeed, by combining HRTEM two-dimensional strain maps of Ge NWs at their base and simulations, Taraci *et al* have clearly shown external loading effects introduced during sample preparation.[33] These examples show that monitoring the strain relaxation at the base of the NWs is a complex task and the first successful attempts have been reported only recently. In 2010, *in-situ* measurements showing the evolution of the in-plane lattice parameter, and thus of the strain state, as the growth time increases have been reported for GaN NWs.[34] However, the observation of pure elastic relaxation is hindered by the plastic relaxation that takes place during the NW nucleation step. In parallel, high-resolution X-ray diffraction has been used to determine the inhomogeneous strain in GaN NWs from the simulation of the associated X-ray peak profiles.[35] Nevertheless, an exponential strain decay had to be assumed, based on previous theoretical calculations,[36] and thus could not be independently confirmed. Only very recently, elastic strain relaxation in top-down fabricated GaN NWs has been experimentally observed for the first time.[37] This work shows that the strain relaxation occurs very rapidly, (i.e. close to the base of the NW) but unfortunately these authors could only observe one intermediate strain state between the initial and final ones. The goals of the work reported here are: first, to determine unambiguously the evolution of the strain-relaxation at the base of the NWs (up to their top), and second, to evaluate how the strain evolves when the NWs are coalesced.

## II.     EXPERIMENTAL PROCEDURE

To monitor the "pure" elastic strain relaxation (i.e. without additional plastic-relaxation) evolution at the NWs base, a top-down approach has been used to fabricate NW array from thick GaN epitaxial layers. The details of the top-down fabrication can be found in references 38 and 39. It should be noted that the fabrication temperature is sufficiently close to room-temperature to neglect dislocation glide. For this study, the etching duration has been varied from 15 to 600 sec, resulting in NWs heights ranging from 30 to 1160 nm and keeping a constant diameter of 265 ± 10 nm (Figure 1). Besides, the fabrication of NWs from GaN layers grown on different substrates (sapphire and silicon) allows study of the relaxation evolution from initially compressively- (on sapphire) or tensilely-strained (on silicon) 1D objects still attached to the starting structure. For the current investigation, the use of our top-down approach has two main advantages: first, the NW height is easily varied (i.e. by changing the etching duration), and second, large samples areas can be studied with negligible size dispersion (*i.e.* negligible inhomogeneous broadening). This allows the use of characterization techniques with large probe areas, like X-ray diffraction and photoluminescence (PL).

Nitride semiconductor materials (GaN, AlN, and InN) possess large piezoelectric coefficients, around ten times larger than for GaAs, and deformation potentials as large as ~ -9 eV for biaxial (0001) strain.[40,41] As a consequence, the band gap energy of nitrides is very sensitive to the strain state. In that sense, compared to more complex structural characterizations, PL is a straightforward way to indirectly determine the strain state via the measurement of the band-gap emission energy. For that purpose, the NWs were excited (at room-temperature) by the 325 nm line of a He-Cd laser and their luminescence was collected by a GaAs multiplier located at the exit of a 64 cm single monochromator.

High-resolution X-ray diffraction experiments were also performed using a standard triple axis Seifert diffractometer equipped with a crystal analyzer. Usually, asymmetric (105) or (205) reflections are used to directly determine the in-plane *a* lattice parameter. Unfortunately, for NWs less than 300 nm high reciprocal space maps around these reflections did not allow to distinguish clearly (i.e. without deconvolution) the signature of the NWs from that of the underlying GaN layer. Therefore, to apply the same methodology to all samples, only 2θ-ω scans around the symmetric (0002), (0004), and (0006) reflections were measured.

## III. RESULTS AND DISCUSSION

### A. Determination of the NWs strain from photoluminescence measurement

Room-temperature PL measurements of GaN NWs with different heights are shown in figure 2 (a). The upper panel shows the emission spectra of NWs of different height fabricated on sapphire whilst the lower shows the emission spectra of the NWs on silicon. The free-exciton emission of the GaN layer on sapphire is centered at 3.426 eV, which is above the commonly reported strain-free value ($\cong 3.41 \pm 0.002$ eV). On the other hand, on silicon the emission occurs at a smaller energy (3.387 eV) than that of strain-free material. The observed energy blue-shift on sapphire is characteristic of GaN under compressive strain (larger bandgap), while the red-shift observed on silicon is related to the presence of tensile strain (smaller bandgap). As the NW height increases (see figure 2b), the free-exciton emission red-shifts on sapphire (red squares), while the opposite behavior is observed for the NWs on silicon (blue circles). To facilitate the observation of the emission energy evolution, the PL intensities have been normalized in figure 2a. Note that, in agreement with theoretical predictions, stronger light absorption and/or extraction for the 1D structure is observed as the NW height increases, leading to a continuous and strong PL intensity enhancement.[39]

For both series, on silicon and sapphire substrates, the evolution of the free-exciton energy as a function of the NW height is shown in Figure 2(b). After a fast energy shift during the first ~200 nm, the free-exciton energy tends towards the same value (3.409 eV), independently of the substrate used. The free-exciton energy at "saturation" is in the same range as the room-temperature values reported for fully-relaxed GaN pillars,[42] and homoepitaxial GaN layers.[43] It should be noted that since the NW diameter is much larger than the Bohr radius of the GaN exciton, no quantum confinement is expected. Therefore, the PL emission of strain-free GaN NWs with the present sizes should match that of bulk strain-free GaN.

When the height of top-down fabricated NWs exceeds 1000 nm (clearly in the saturation regime), several sharp peaks appear on the low-temperature PL spectra (not shown here but reported in reference 39]. These emission lines at 3.471, 3.478, and 3.483 eV correspond to the well-known GaN donor-bound exciton, and free A- and B- excitons, respectively. Their energy positions are in perfect agreement with those already reported for bottom-up grown GaN NWs,[44,45] for GaN NWs detached from the substrate,[46,47] for strain-free GaN quasi-substrates,[48,49] and for GaN homoepitaxial layers on bulk crystal substrates.[43,50] All these observations indicate that NWs are strain-free once their height reaches 300 nm (i.e. when their aspect ratio height/diameter ~ 1).

As predicted by deformation potential theory,[51] the band-gap energy for most semiconductor materials changes when they are subject to biaxial strain. Owing to the possibility of growing GaN on various substrates like sapphire, silicon, or silicon-carbide, the linear relationship between the A free-exciton emission energy and the in-plane strain has been demonstrated for a large range of tensile and compressive strains.[40] The same linear behavior has also been demonstrated from mechanically bent ZnO NWs.[52] Using the free-exciton energy $\left(E_{FE}^{strain-free}\right)$ extracted from the room-temperature PL measurements, the biaxial in-plane strain ($\varepsilon_{xx}$) can been calculated using the following linear relation:

$$E_{FE} = E_{FE}^{strain-free} + \frac{\partial E_{FE}}{\partial \varepsilon_{xx}} \times \varepsilon_{xx} \qquad (1)$$

where $E_{FE}^{strain-free}$ is the strain-free free-exciton energy, and $\frac{\partial E_{FE}}{\partial \varepsilon_{xx}}$ corresponds to the GaN deformation potential. Using the $E_{FE}^{strain-free}$ value determined above (3.409 eV) and the well accepted GaN potential deformation value of -9 ± 1 eV,[40,41] the in-plane strain evolution has been extracted as a function of the NWs height. The strain profile determined by this procedure is presented in section C, where it will be compared with that extracted from the X-ray diffraction measurements (detailed in the next section) and the results of 3D strain calculations.

While the focus of this article is the strain-relaxation within the NWs and, therefore, from the optical point of view we are mostly interested in the energy of free-excitons, it should be noted that the PL spectra in figures 2 (a) and (b) show a number of additional peaks at lower energies. If we concentrate on the GaN NWs on sapphire (unetched sample) , where the peaks are more visible, we can first identify the energy band centered around 3.36 eV with the first LO phonon replica, since its energy separation from the free-exciton emission equals $E_{LO} - 3/2 \, (kT)$, $E_{LO}$ being the energy of the longitudinal optical phonon. Furthermore, this emission is modulated by several additional peaks certainly due to Fabry-Perot interferences in the cavity formed by the residual planar GaN layer and the sapphire substrate. These interference fringes progressively disappear as the surface becomes rougher (NW height increases). However, the PL spectrum of 300 nm high NWs clearly presents two additional peaks at 3.358 eV and 3.379 eV, corresponding to the LO phonon replica of the free-exciton (A- and B-excitons) and the C-exciton, respectively. This is in agreement with the fact that the lateral free-surfaces of the NWs allows for the observation of the C-exciton emission (not shown here, but visible in logarithm scale). Finally, the reasons why the intensity of the phonon replica increases as the NW height increases and why they are not so clearly observed for NWs on silicon are still unsolved questions, but their study is clearly out of the scope of this work.

### B. Determination of the NW strain from X-ray diffraction measurements

The 2θ-ω scans on the (0004) reflection for the NWs series on sapphire are shown in Figure 3 (a). For the initial template, only one peak is present at smaller angles than the strain-free bulk GaN. This observation confirms that the initial GaN template on sapphire is under compressive biaxial strain. As the height of the NWs increases by progressively etching into the 2D layer, a second peak at higher 2θ angles appears. Moreover, as the etching duration increases, the intensity of the small angle peak drops while the intensity of the second peak increases continuously. Finally, when the NWs are etched completely through the 2D layer (spectrum not shown on figure 3), only the second peak is present. This confirms that the second peak is related to the NWs.

A fitting procedure based on pseudo-Voigt functions has been carried out to extract the line-width (FWHM) and the position of the 2D layer and NW peaks. The FWHMs of the two peaks have the opposite behaviour [Figure 3 (b) – top panel]. As the etching duration increases, the FWHM for the 2D layer the FWHM continuously increases whilst that of the NW peak decreases. For 2θ-ω scans, the peak profile can be influenced by several effects, the most significant being the presence of strain gradients or strain inhomogeneities within the investigated volume.[35] A number of studies have shown that the strain inside GaN layers is not homogeneously distributed across the entire thickness.[53-55] Indeed, the strain is maximum at the interface with the substrate and exponentially decreases as the GaN layer thickness increases.[53-55] Thus, in our case as the etching duration increases the 2D layer gets thinner and, therefore, the region where the strain gradient is largest contributes more to the low angle peak profile. Consequently, the x-ray diffraction peak is broadened as seen in figure 3. The same argument can also be used to explain the opposite trend observed for the NW peak FWHM. Indeed, several theoretical works predict an exponential strain decay.[36] In that case, the strain gradient should be maximum close to the base of the NW (i.e. at the interface with the underneath 2D GaN layer) and rapidly decreases as the NW height increases, thus explaining the observed decrease of the NW peak FWHM.

The second information extracted from the 2θ-ω scans is the peak positions as a function of etched depth [Figure 3 (b) – bottom panel]. While those of the 2D layer remain nearly constant, the peak corresponding to the NWs rapidly shifts to larger angles before saturating as the NW height exceeds 300 nm. Using the 2θ peak positions, the corrected $c$ lattice parameter has been directly calculated using the following equation:

$$c = l \times \frac{\lambda}{2 \times \sin(\theta)} \times \left(1 + \frac{\delta}{\sin(2\theta)}\right) \qquad (2)$$

with $\lambda$ corresponding to the X-ray radiation wavelength (1.54056 Å), $\delta$ being unity minus the refractive index ($1.73 \times 10^{-5}$ $e/\text{Å}$, for GaN),[56] and $l$ being the number corresponding to the $hkl$ atomic plane probed. The red squares in figure 3 (c) correspond to the $c$ lattice parameters determined from the NW series on sapphire. To reduce the measurement inaccuracy, each square corresponds to the average value of the $c$ lattice parameters extracted from the (002) and (004) reflections. Using the same procedure, the $c$ lattice parameters have been also determined for NWs with different heights on a silicon substrate and are shown by blue circles in Figure 3 (c). As a result of different initial strain states and, thus, opposite $c$ lattice parameter evolution, the general behaviors of the two NWs series are the mirror image of each other. Further, after a fast variation during the first 200 nm (decrease of the c lattice parameter for the NWs on sapphire and increase for the ones on silicon), the extracted $c$ lattice parameter values remain constant as the NW height exceeds 250 - 300 nm [see insert in Figure 3 (c)]. For both series, the $c$ lattice parameter saturates at 5.185 Å. For many years, large discrepancies existed for the strain-free lattice parameters and values for the Poisson's ratios for III-nitrides. However, the continuous improvement of the material quality (i.e defect density and impurity concentration reductions), and the refinement of the methods used to determine these parameters, have permitted to extract universal values. For GaN, the strain-free $a_0$ and $c_0$ lattice parameters are 3.189 Å and 5.185 Å, respectively.[56-58] By comparison with the strain-free $c_0$ lattice parameter value, the saturation observed in figure 3 (c) can be assigned to the fully-relaxed condition.

Assuming a biaxial strain, consistent with the symmetry of our NWs (cylindrical) and of the employed substrates, the $c$ lattice parameter reported in figure 3 (c) can be used to extract the in-plane strain using the equation:

$$\varepsilon_{xx} = \frac{c-c_0}{c_0} \times \frac{\nu-1}{2\nu} \quad (3)$$

where $c_0$ is the strain-free lattice parameter and $\nu$ is the Poisson ratio. Despite the hexagonal wurtzite structure, hydrostatic pressure experiments on GaN have highlighted an isotropic behavior,[59] so a single Poisson ratio of 0.183 has been used.[56] The in-plane strain profile determined from equation (3) is shown in figure 4 (b) and is discussed in the next section.

### C. Comparison of experimental and calculated strain profiles along the NW height

The in-plane strain ($\varepsilon_{xx}$) profiles extracted from the PL measurements are reported as a function of the NW height in figure 4 (a). The red squares correspond to the GaN NWs on sapphire substrates while the blue circles correspond to the NW on silicon (open and filled circles represent NW series with different initial strains). As expected from the variation of the free-exciton energy [figure 2 (b)], the in-plane strain rapidly decreases during the first 100-200 nm and, independently of the initial strain nature (tensile or compressive), the NWs are almost fully-relaxed when their height reach 300 nm. To validate the in-plane strain profiles deduced from PL, the $\varepsilon_{xx}$ profiles extracted from the X-ray diffraction experiments have been reported in figure 4 (b). In addition to showing similar behaviors, the in-plane strain values and the NW height for which they can be considered fully-relaxed agree closely with the values determined from the analysis of the PL measurements.

At this stage it is important to note that the penetration depth in GaN of the laser used to excite the PL is 150-200 nm, while it amounts to several micrometers for X-rays. The in-plane strain thus determined for each NW height is an averaged value along their length and not a local value. This means that the strain relaxation rate is underestimated. However, the error associated with its determination is small because the relaxation process is abrupt. Indeed, both methods indicate an exponential decay of the strain along the NW height, in agreement with theoretical predictions,[36] and recent indirect,[35] and direct,[37] experimental demonstrations. The remarkable similarity between the evolution of the relaxation with NW height on both sapphire and silicon substrates, despite the opposite sign of the initial strains, is certainly due to the combination of identical NW diameters and similar absolute values for the initial strains. X-ray diffraction measurements of the initial 2D layers indicate that these amount to 0.21 % for GaN on silicon and 0.18 % for GaN on sapphire. Comparable initial strain amounts were confirmed by the PL measurements. Indeed, the emission energy of the 2D GaN layer on silicon is red-shifted by 22 meV (with respect to the strain-free emission energy) compared to a blue-shift of 18 meV for the layer on Sapphire [Figure 2 (b)].

To confirm the quantitative strain relaxation behavior deduced from the PL and the X-ray diffraction measurements, 3D strain calculations based on a continuous media approach have been carried out using the multiscale software tool TiberCAD.[60,61] In TiberCAD, the calculation of strain in lattice mismatched heterostructures is based on the linear elasticity theory of solids, assuming pseudomorphic interfaces between different materials.[62] Small deformations are assumed, so that strain can be considered a linear function of deformation and Hooke's law can be used to relate stress to strain. The strain and deformation fields are found by minimizing the elastic energy of the system. As a result, the strain tensor in any point of the structure is obtained. For the current calculations, the strain has been assumed homogeneous in the initial template and its value has been imposed to be the one determined by X-ray diffraction measurements. A vanishing displacement in the direction normal to the lateral template surface has been imposed to account for an infinite extension of the material in the lateral direction. The average strain evolution as a function of the GaN NW height has been calculated by integrating the strain in the NW. The 3D strain repartition inside the NW has been calculated for several NW heights, ranging from 25 to 1200 nm and a fixed 250 nm diameter.

The maps of the in-plane strain calculated for NWs on sapphire with heights of 25, 75, 150, and 300 nm are shown in figure 4 (c). For the smallest height, the compressive strain spreads inside the whole NW. As the NW height increases to 75 nm, it becomes clear that the strain remains confined in a region localized at the base and in the central part of the NW, while the top rims are almost strain-free. This strain repartition is even clearer for the 300 nm high NW for which all the strain is located at the base while most of the top part is strain-free. This predicted strain distribution supports the view that the strain relaxation rate extracted from the PL and X-ray diffraction measurements is underestimated.

In order to compare the calculation results with the strain determined from the X-ray diffraction measurements, an averaged strain for a given height has been obtained by summing the strain of each pixel forming the NW, and dividing by its volume. The calculated average strains, represented as black lines in figure 4 (a) and (b), closely match the experimental results. Since the calculations do not use any fitting parameter, the almost-perfect agreement obtained confirms the validity of the experimental procedures and of the use of a continuous media calculation method to apprehend elastic relaxation in such NW structures.

### D. Strain state inside coalesced NWs template

One of the advantages often put forward to promote the use of GaN NWs as passive templates, as described in the introduction, is the possibility that the subsequent coalesced layer could be strain-free.[63] Based on the study of strain relaxation described above (sections A, B, and C), an array of 1000 nm high GaN NWs on a silicon substrate was fabricated to ensure an initial strain-free state. Then, the coalescence of the passive NWs template has been done using Metal Organic Vapor Phase Epitaxy (MOVPE) in a close-coupled showerhead reactor. The details of the coalescence process will be reported elsewhere. The final coalesced GaN layer (around 1µm thick from cross-section scanning electron microscopy) presents a very smooth surface (without any hole or pit) with a root-mean-square roughness below 1nm, obtained from 5×5 µm² atomic force microscopy images.

In order to determine whether or not the coalesced GaN layer is strain-free, room-temperature PL and X-ray diffraction measurements have been carried out. The PL emission of the coalesced template is clearly red-shifted compared to that of the initial NWs template [Figure 5 (a)]. Indeed, the emission energy of the GaN layer is 3.400 eV, while the NWs emission is centered at 3.409 eV. Since the coalesced layer is nominally undoped, this emission shift is a clear indication of an accumulated tensile strain. Using equation (1), an in-plane strain $\varepsilon_{xx}$ of 0.10 % is deduced. The X-ray diffraction measurement confirms the tensile strain state of the coalesced layer, since the peak appears at larger angles than that of the strain-free NWs [figure 5 (b)]. Interestingly, on the 2θ-ω (004) scan of the coalesced layer, a small bump around 72.9° is also visible. The large penetration depth of the X-rays allows probing of the complete structure, suggesting that the observed bump is a signature of the buried NWs. In that case, this would indicate that the NWs remain almost strain-free after the coalescence. It is important to state that the overgrowth process used, allows obtaining a fully coalesced layer while keeping buried air gaps between the NWs (see Figure 5(c)). Using the method detailed in section III. B., an in-plane strain of 0.13 % is extracted from the (004) scan of the coalesced layer. The in-plane strain obtained from the PL and X-ray diffraction measurements are not perfectly identical. Since the probing depths of the two techniques are very different (around 200 nm for PL versus several micrometers for X-ray), the obtained disparity might result from a strain gradient across the coalesced layer thickness.

As already demonstrated in thick GaN layers,[53] laterally overgrown GaN,[64,65] InGaN/GaN micro-pillars,[66] and coalesced GaN NWs,[25] cathodoluminescence (CL) spectroscopy is a powerful technique to visualize strain gradients, given its spatial and spectral resolutions. For mapping the strain evolution after coalescence, the sample was prepared for cross-section observation. CL experiments have been realized in a JEOL JSM 7000f scanning electron microscope (SEM) using an acceleration voltage of 5 kV. To allow a direct comparison with the PL and X-ray diffraction results, the CL measurement has been carried out at room-temperature. The cross-section SEM image of the complete coalesced structure is reported in figure 5 (c). The "dark" grey part on the left side corresponds to the silicon substrate, while the sample surface is at the right of the image. Initially, the GaN NWs height was roughly 1000 nm and they were separated 600 nm from each other. The periodic dark regions indicate that air gaps are still present after the coalescence, although a large part of the initial gaps has been filled by GaN during the overgrowth step.

In Figure 5(c) the red-dashed box defines the selected area (2.35 μm x 1 μm) in which CL mapping has been performed using a pixel size of 100×100 nm$^2$. For each pixel, the emission spectrum has been fitted using a non-linear least squares algorithm, to accurately extract the near band edge emission energy. In order to present the data, a method usually applied to extract quantitative strain maps from high-resolution transmission electron microscopy has been used. In brief, the emission energy of a given region is taken as a reference, and the strain difference map is constructed by first calculating at each pixel the ratio ($E_{\text{pixel}}$ − $E_{\text{reference}}$) / $E_{\text{reference}}$, and then by applying the relationship between the emission energy and the strain used in section III.A. The parts between the air gaps, corresponding to the initial NWs, have been chosen as reference and the $\Delta\boldsymbol{\varepsilon_{xx}}$ map has been determined [Figure 5 (d)]. The black squares correspond to the air gap positions. The $\Delta\boldsymbol{\varepsilon_{xx}}$ map clearly shows that the GaN coalesced layer just on top of the NWs is under tensile strain (it should be recalled that the substrate is silicon). While the strain distribution is quite inhomogeneous, its value oscillates between 0.07 and 0.13 %, in agreement with the PL and X-ray diffraction measurement results. Strong strain inhomogeneities in coalesced GaN layers has been also reported by Lethy et al.[25] The main observation, based on three different techniques (PL, X-ray diffraction and CL), is that strain is unambiguously rebuilt during the coalescence of strain-free NWs, confirming some of the previous studies.[23,25]

**CONCLUSION**

The strain relaxation rate at the base of NWs (250 nm in diameter) still in contact with the substrate has been measured. "Pure" elastic strain relaxation has been independently extracted from PL and X-ray diffraction measurements in two series (on sapphire and silicon) of NWs whose height has been continuously varied. Both techniques provide evidence of a very fast strain relaxation within the first one to two hundreds nanometers. This strain evolution along the nanowire height is in good agreement with the results of simulation based on a continuous media approach. This experimental observation of the strain evolution constitutes a step forward towards the understanding of strain relaxation in 1D systems, and provides practical guidance for the design of active-nanowire based devices and passive-nanowire based templates. The study of coalesced NW arrays has highlighted a strain rebuilt process. The next step will be to determine if the accumulated strain is smaller than that of comparable conventional structures, and how it evolves as the coalesced layer thickness increases.

**Acknowledgements**

This work was supported by the European Union under FP7 contract SMASH CP-IP- 228999-2. The authors are grateful to Dr Mathieu Leroux for manuscript reading and fruitful discussions.

**FIGURE CAPTIONS**

Figure 1. Scanning electron microscopy images of top-down fabricated GaN nanowires for etching duration of 60 sec (a), 150 sec (b), 300 sec (c), and 600 sec (d). The sample inclination angle is 20°.

Figure 2. Normalized room-temperature photoluminescence spectra of unetched GaN layer and 60 nm, 175 nm, and 300 nm high GaN nanowires (a), on sapphire (upper panel) and on silicon (bottom panel) substrates. Evolution of the band-edge emission energy as a function of the nanowire height (b). The red squares correspond to the nanowires series on sapphire substrate while the filled and the opened blues circles correspond to two different nanowires series on silicon substrates (color online).

Figure 3. X-ray diffraction 2θ-ω scans of unetched GaN layer and 120 nm, 300 nm, 600 nm and 1200 nm high GaN nanowires on sapphire substrate (a). Full-width-at-half-maximum (upper panel) and position (bottom panel) of GaN nanowires and GaN layer peaks (b). Evolution of the $c$ lattice parameter as a function of the nanowire height (c). The red squares correspond to the series on sapphire substrate while the blue circles correspond to the series on silicon substrate (color online). The insert shows the $c$ lattice parameter variation across the first 300 nm from the nanowire base.

Figure 4. In-plane strain ($\varepsilon_{xx}$) evolutions deduced from the PL measurements (a) and the X-ray diffraction experiments (b). The red squares correspond to the nanowire series on sapphire substrate while the filled and the open blue circles correspond to two different nanowire series on silicon substrate (color online). The black lines are the calculated averaged in-plane strain ($\overline{\varepsilon_{xx}}$) from 3D strain distribution simulations. Simulated strain maps of 25 nm, 75 nm, 150 nm, and 300 nm high GaN nanowires on top of a GaN layer on sapphire substrate (c).

Figure 5. Normalized room-temperature PL spectra (a), and X-ray diffraction 2θ-ω scans (b), of strain-free GaN nanowires (blue line) and coalesced layer (red line). Cross-section scanning electron microscopy image of the complete coalesced structure (c). The "dark" grey part on the left side corresponds to the silicon substrate, while the sample surface is at the right of the image. The red-dashed box defines the selected area (2.35 µm x 1 µm) in which cathodoluminescence mapping has been performed. Strain difference map ($\Delta\varepsilon_{xx}$) deduced from the cathodoluminescence mapping measurement (d).

**FIGURES**

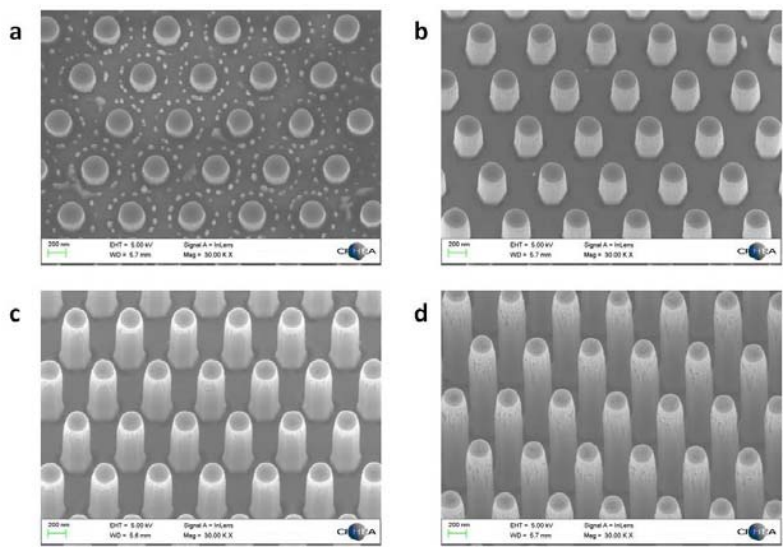

**Figure 1**

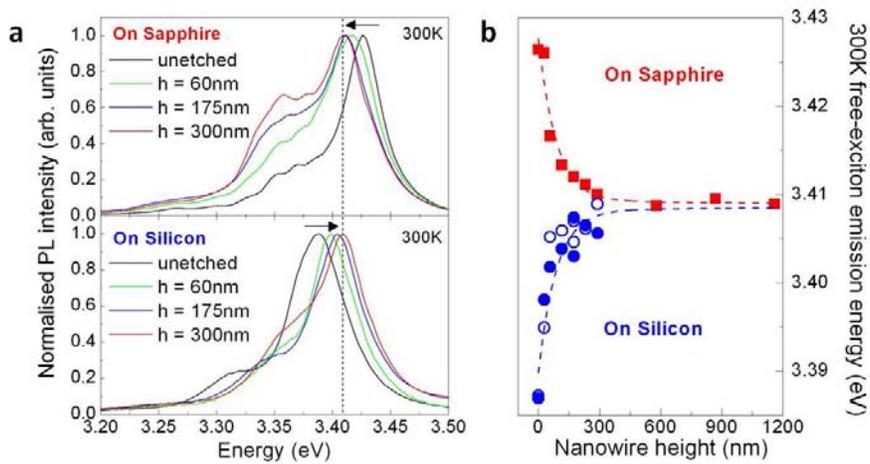

**Figure 2**

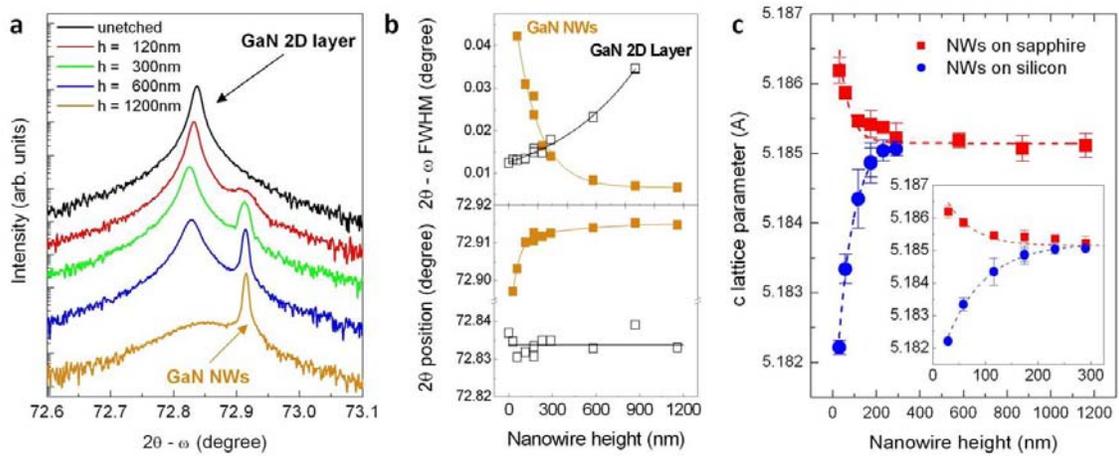

**Figure 3**

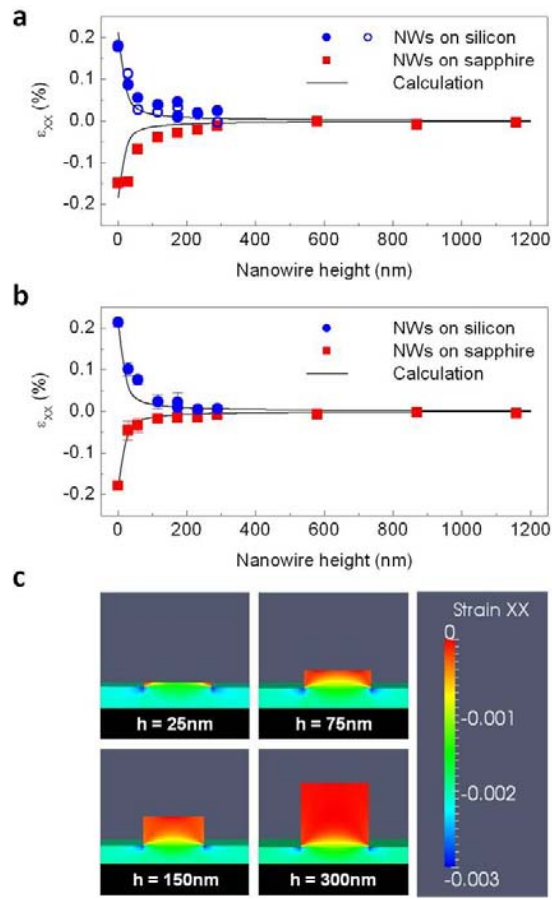

**Figure 4**

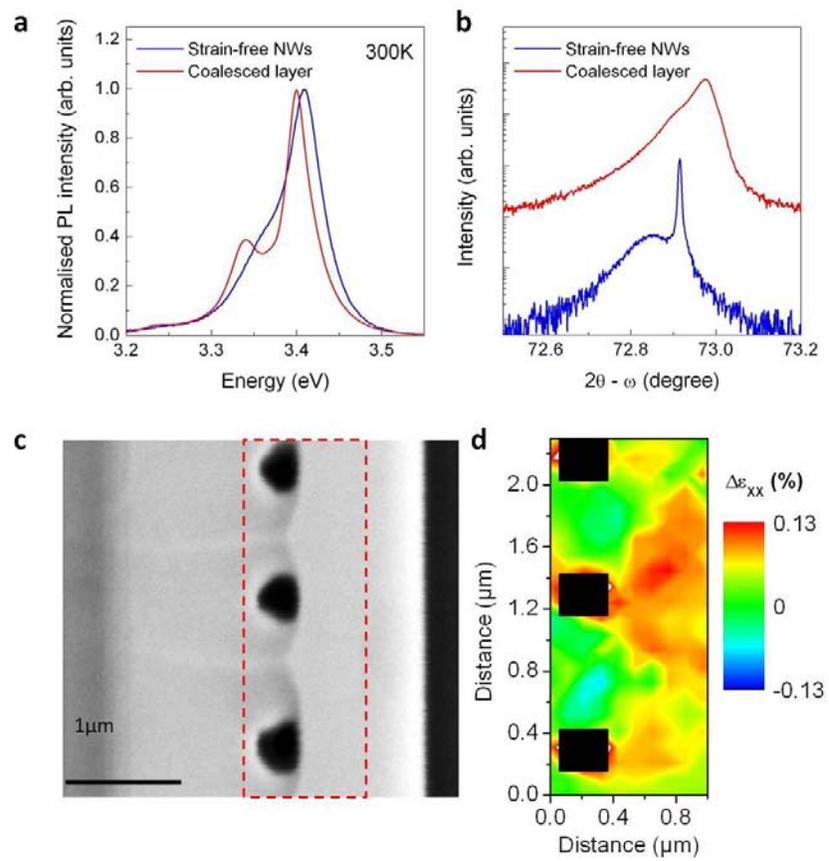

**Figure 5**